\documentclass[conference]{IEEEtran}
\usepackage{graphicx,epsfig,amsmath,amssymb,bm}
\usepackage{amssymb,balance}
\usepackage{amsmath}
\usepackage{amsthm} 
\usepackage{amssymb}
\usepackage{algorithmicx}
\usepackage{algpseudocode}
\usepackage{cite}
\usepackage{url}
\newfont{\bbb}{msbm10 scaled 500}

\newfont{\bb}{msbm10 scaled 1100}

\usepackage{bbm}
\usepackage{subcaption}
\usepackage{tabularx}
\usepackage{mathtools}
\usepackage{empheq}
\usepackage{multirow}
\usepackage{verbatim}
\usepackage[ruled,linesnumbered]{algorithm2e}
\usepackage{color}
\usepackage{tikz}
\usepackage[acronym]{glossaries}
\usepackage[capitalise,noabbrev]{cleveref}
\include{defn}

\newcommand{\beqa}{\begin{eqnarray}}
\newcommand{\eeqa}{\end{eqnarray}}

\ifCLASSINFOpdf

\else

\fi

\begin{document}

\title{Analysis of Load-Altering Attacks Against Power Grids: A Rare-Event Sampling Approach \vspace{-0.2 cm}}
\IEEEoverridecommandlockouts 
\author{
\IEEEauthorblockN{Maldon Patrice Goodridge\IEEEauthorrefmark{1}, Subhash Lakshminarayana\IEEEauthorrefmark{2} and Christopher Few\IEEEauthorrefmark{3} } 
\IEEEauthorblockA{\IEEEauthorrefmark{1}Global Development Initiatives, Queen Mary University of London \\
\IEEEauthorrefmark{2}School of Engineering, University of Warwick, UK \\
\IEEEauthorrefmark{3}National Grid, UK \\
Emails: \IEEEauthorrefmark{1}m.p.goodridge@qmul.ac.uk, \IEEEauthorrefmark{2}subhash.lakshminarayana@warwick.ac.uk,
\IEEEauthorrefmark{3}christopher.few@nationalgrid.com
}

\vspace*{-0.8 cm}
}



\maketitle

\begin{abstract}
By manipulating tens of thousands of internet-of-things (IoT) enabled high-wattage electrical appliances (e.g., WiFi-controlled air-conditioners), large-scale load-altering attacks (LAAs) can cause severe disruptions to power grid operations. In this work, we present a \emph{rare-event sampling} approach to identify LAAs that lead to critical network failure events (defined by the activation of a power grid emergency response (ER)). The proposed sampler is designed to ‘skip’ over LAA instances that are of little interest (i.e., those that do not trigger network failure), thus significantly reducing the computational complexity in identifying the impactful LAAs. We perform extensive simulations of LAAs using the Kundur two-area system (KTAS) power network while employing the rare-event sampler. The results help us identify the victim nodes from which the attacker can launch the most impactful attacks and provide insights into how the spatial distribution of LAAs triggers the activation of ERs.
\end{abstract}

\IEEEpeerreviewmaketitle

\section{Introduction}
Cyber attacks against power system grids can have significant social and economic consequences. The threats can be broadly divided into two categories -- (i) attacks that directly target the power grid's supervisory control and data acquisition (SCADA) system, and (ii) attacks that indirectly target the power grid's control loops by manipulating end-user internet-of-things (IoT) enabled electrical appliances. Direct attacks against the SCADA system, such as false data injection attacks and/or coordinated cyber-physical attacks against power grid state estimation have received significant attention  \cite{Liu2009, LakshGT2021, LakshRMT2021}. In contrast, indirect attacks that target a large number of demand-side appliances in a Botnet-type attack have been studied only recently \cite{Dabrowski2017, Soltan2018}. Unlike the SCADA assets, these devices cannot be monitored continuously due to their large numbers. 



The focus of this work is on 
\emph{load-altering attacks} (LAAs), which refer to a sudden and abrupt change in the power grid demand by synchronously turning on/off a large number of IoT-enabled high-wattage appliances \cite{Dabrowski2017, Soltan2018, HuangUSENIX2019, AminiLAA2018, LakshIoT2021}. LAAs pose a major threat to power grid operations since they can potentially disrupt the balance between supply and demand. It has been shown \cite{Dabrowski2017, Soltan2018} that such attacks can lead to unsafe frequency excursions, line outages, and/or increase  the grid's operational costs. Moreover, dynamic LAAs, in which the attacker injects a series of load perturbations over time, can also destabilize the power grid's frequency control loop \cite{AminiLAA2018}. Subsequent work has also focused on detecting LAAs using data-driven approaches based on the data gathered from phasor measurement units \cite{AminiIdentification2019, lakshminarayana2021datadriven}.

Understanding the impact of LAAs is an important component in risk analysis. Existing work on quantifying the impact of LAAs can be categorized into two approaches -- (i) simulation-based  approach \cite{Dabrowski2017, Soltan2018, HuangUSENIX2019, AminiLAA2018}, and (ii) analytical approach \cite{LakshIoT2021}. Under the former, the attack impact is assessed by simulating the power grid's control loops (e.g., frequency dynamics) \cite{Dabrowski2017, Soltan2018, HuangUSENIX2019} or computing the system's eigenvalues under LAAs \cite{AminiLAA2018}. However, the results presented in these works correspond to only a few specific LAA scenarios (i.e., specific load perturbations injected at the victim nodes). Evaluating the attack impact under all possible spatial distributions of LAAs (over the victim nodes) requires performing extensive simulations considering different combinations of the victim nodes and attack magnitudes, which can be computationally prohibitive. To overcome these issues, an analytical approach based on the theory of second-order dynamical systems was proposed in \cite{LakshIoT2021}. The closed-form analytical functions to evaluate the attack impact (in terms of the dynamic response and eigensolutions of the power grid's frequency control loop) only need to be computed once, thus avoiding the requirement for repeated simulations. However, the analytical approach is restricted to a second-order model with a direct current (DC) power flow model. Extending these results to higher-order models (e.g., one that considers both frequency and voltage dynamics) and 
involving the non-linear alternating current (AC) power flow model is non-trivial. This is important to obtain a realistic assessment of the LAA impact.



To overcome the aforementioned limitations, we apply a sampling approach to map LAA magnitudes and their spatial distributions (across the different victim nodes) to their attack impact. In particular, the focus of this work is on LAAs that lead to the activation of power system emergency responses (ERs) which disconnect critical power system components (e.g., generators/load/transmission lines). A common sampling approach is \emph{Monte Carlo} simulations, which would apply randomly generated realisations of LAAs from an underlying distribution to a simulated model of the physical system. However, power grid design philosophies, such as $N-1$ scheduling, make the grid resilient to various contingencies (including cyber attacks). This implies component disconnections induced by LAAs can be extremely rare. Consequently, Monte Carlo sampling can be computationally expensive, as the majority of sampled LAAs will not result in component disconnections, requiring a large number of realisations applied to the power system model to generate a sample of the rare event (i.e., the activation of an ER).

To avoid an exhaustive search to locate potential LAAs that lead to network failures, we instead employ a novel methodology in the context of LAAs based on a Markov chain Monte Carlo (MCMC) approach for rare-event sampling, known as the \emph{skipping sampler}. The proposed approach is designed to reduce the time and computational effort spent on evaluating LAAs that are of little interest (i.e. those that do not result in an emergency response), allowing it to more efficiently construct a sample of a rare event. The skipping sampler has been applied to draw samples of low probability, high impact events in power networks in the literature (see [12] and [13]). Section~\ref{sc:skipping_sampler} provides a detailed discussion on the skipping sampler.

We evaluate the framework by performing extensive simulations using the Kundur two-area system (KTAS) \cite{kundur1994power}. We simulate the power grid's transient dynamics by a third-order model, which accounts for both the frequency as well as the voltage dynamics \cite{schm2014}. The LAAs that 
perturb these dynamics are modelled according to the \emph{Log-normal} distribution (since real-world cyber attack magnitudes are well modelled by this distribution \cite{Edwards2016}). If the local frequency metrics at any node exceed pre-set tolerances, appropriate ERs, such as generation/load shedding, are activated. Our results show that in the KTAS network, most instances of ERs are activated for two specific spatial distributions of LAAs, namely (i) when the attacker increases the load throughout the system or (ii) when the attacker decreases the load in the over-provisioned area (with excess generation) and increases the load in the under-provisioned area (with excess load). In particular, attacks that exacerbate the power imbalance in the system (i.e., type (ii)) can trigger inter-connector line disconnections, and lead to other network failures.

The rest of the paper is organised as follows. Section \ref{sec:Prelim} introduces the system model; Section~\ref{sc:statmodel} presents the statistical model and details of the rare event sampling approach.
Section~\ref{sec:Sims} describes the simulation results and Section~\ref{sec:Conc} concludes. The simulation parameters are provided in an online Appendix found in \cite{extended}.

\section{System model}\label{sec:Prelim}
\subsection{Power Grid Model}

Using the third-order model for the generator, the power system model for rare-event sampling simulations will also include a model for governor action, automatic voltage regulation and a model of protection system ERs if the frequency or RoCoF exceeds pre-defined thresholds. We consider a power grid represented by $\mathcal{G} = \{ \mathcal{N}, \mathcal{W} \},$ where $\mathcal{N}$ is the set of buses and $\mathcal{W}$ is the set of transmission lines. The set of buses $\mathcal{N}$ consists of $N$ generation buses and $L$ load buses, with $|\mathcal{N}| = N+L$. At each generation bus $i= 1, \dots N$, the dynamics for the phase angle $\delta_i$, voltage magnitude $E_i$ and governor action $\rho_i$ are given respectively by:

{\small\begin{subequations}
	\label{eq:network_model}
	\begin{empheq}[left = \empheqlbrace\,]{flalign}
			& M(\psi) \ddot{\delta}_{i}+D \dot{\delta}_{i}=\psi_i \chi^G_i- \chi^L_i(\mathcal{R}_i)-\nonumber \\ &\hspace{3cm}E_{i}\sum_{j=1}^{N+L}B_{ij}(\Omega_{ij})E_{j}\sin(\delta_{ij} )\label{eq:freq21} \\
			& S_{i}\dot{E}_{i}=\psi_i (E_{f,i}-\text{v}_i) -E_{i}+ X_{i}\sum_{j=1}^{N+L}B_{ij}(\Omega_{ij})E_{j}\cos(\delta_{ij} )\label{eq:volt21}\\
			& \dot{\rho}_{i}= -A_{i}\dot{\delta}_{i}(1-1_{\mathcal{\mathcal{\mathcal{W}}}}[\dot{\delta}_{i}]).\label{eq:gov21}
		\end{empheq}
	\end{subequations}}
	
\noindent In a similar manner, the dynamics for $\delta_i$ and $E_i$ at each load bus $i=N+1, \dots, N+L$ are given by: 
{\small \begin{subequations}
	\label{eq:network_model_load}
	\begin{empheq}[left = \empheqlbrace\,]{flalign}
			& M(\psi) \ddot{\delta}_{i}+D \dot{\delta}_{i}= -\chi^L_i(\mathcal{R}_i)- E_{i}\sum_{j=1}^{N+L}B_{ij}(\Omega_{ij})E_{j}\sin(\delta_{ij})\label{eq:freq_load} \\
			& S_{i}\dot{E}_{i}=\psi_i E_{f,i}-E_{i}+X_{i}\sum_{j=1}^{N+L}B_{ij}(\Omega_{ij})E_{j}\cos(\delta_{ij}. )\label{eq:volt_load}
    \end{empheq}
\end{subequations}}
In equations~\eqref{eq:network_model} and ~\eqref{eq:network_model_load}, $\psi_i$, $\Omega_{ij}$ and $\mathcal{R}_i$ are indicator variables associated with generator, line and load disconnections respectively, explained in Section~\ref{sec:emer_res}. The system angular momentum, $M(\psi)= \sum_{j=1}^{N} \psi_j H_j$ is given by the sum of each generator's inertia constant, $H_i$ (see online Appendix for parameter values).
 
\begin{table}[!h]
    \centering
    \caption{Variables used in ~\eqref{eq:network_model} and ~\eqref{eq:network_model_load}. }
    \begin{tabular}{|c|l|c|}
    \hline
        \textbf{Symbol}&\textbf{Meaning}&\textbf{Units}\\\hline
         $A_i$&Governor's droop response &MW/rad\\                 $B_{ij}(\Omega_{ij})$&Susceptance matrix&p.u.\\   $\chi_i^G$&Net generation at node $i$&p.u.\\
         $\chi_i^L(\mathcal{R})$&Net loads at node $i$&p.u.\\
         $D$&System damping& \%\\
         $\delta_i$ & Phase angle & p.u\\
         $\delta_{ij}$&$\delta_i - \delta_j$&p.u.\\
         $\dot{\delta}_i$&Frequency & p.u\\
         $\ddot{\delta}_i$&Rate of change of frequency (RoCoF)&p.u.\\
         $E_i$ & Voltage&p.u.\\ 
         $E_{f,i}$&Machine $i$ rotor field voltage&p.u.\\         $M(\psi)$&System angular momentum & Ws$^2$\\
         $\Omega_{ij}$&Line disconnection indicator&-\\
         $\psi_i$ & Generator shed indicator&-\\
         $\mathcal{R}_i$&UFLS counter&-\\
         $S_i$&Machine $i$ transient time constant&s\\
         $X_i$&Machine $i$ equivalent reactance&ohms\\
         $\mathcal{W}$&Governor's deadband frequency range&Hz\\

         \hline
\end{tabular}
\label{tb:model_variables}
\end{table}
As we assume the network to be lossless, the elements of $B_{ij}(\Omega_{ij})$ correspond to the imaginary part of the elements of the network's admittance matrix \cite{kundur1994power}. The net generation at node $i$ is given by $\chi^G_i = \min\{P_i^{\text{max}},P^G_i + \rho_i\}$, where $P_i^{\text{max}}$ is the nominal maximum power output of generator $i$, $P_i^G$ is the equilibrium power of the generator and $\rho_i$ is the power contributed by a governor unit, whose dynamics are given in \eqref{eq:gov21}. The variable v$_i$ accounts for the action of automatic voltage regulation (see online Appendix). The net load at node $i$, $\chi^L_i$, is inclusive of the LAA and a load disconnection scheme, and is discussed in Sections~\ref{sec: laa_model} and \ref{sec:emer_res}. The remaining parameters are given in Table~\ref{tb:model_variables}.
 
 \subsection{Load-Altering Attack Model}
 \label{sec: laa_model}
Several security vulnerabilities have been identified in IoT-enabled high-wattage consumer appliances (see, e.g., \cite{Soltan2018}). These vulnerabilities can be exploited by a strategic attacker to cause security incidents such as information disclosure and privilege  escalation, leading to a change in the device's operational settings (e.g., switch ON/OFF or change the mode of operation). Considering a $2~$kW power rating for the ACs, and a Botnet-scale attack that potentially  compromises tens of thousands of such devices, LAAs can lead to a sudden load change of several MWs of power \cite{Soltan2018}.




Not all loads are expected to be susceptible to LAAs, thus we decompose $P^L_i$, the equilibrium load at bus~$i$, into a vulnerable part, given by $\nu P^L_i$, where $\nu \in [0,1]$ denotes the proportion of equilibrium loads in the network vulnerable to an LAA; and a secure part (i.e., protected or non-smart loads) $(1-\nu)P^L_i$. LAAs at node $i$, denoted $u_i$, are modelled as $u_i \coloneqq \eta_i\nu P^L_i$, where $\eta_i \in [-1,1]$ is the proportional change to equilibrium vulnerable load. Thus, the load at node $i$, inclusive of the LAA, is given by
{ \begin{align}
    \chi^L_i &=  (1-\nu)P^L_i + \nu P^L_i + u_i \nonumber\\ 
    & = (1-\nu)P^L_i +  (1+\eta_i)\nu P^L_i.
\end{align}}
This constrains the authority of the attacker to alter vulnerable loads at node $i$ to a minimum of $0$~MW (i.e. $\min(u_i) = -P^L_i \nu$), or, at maximum, double vulnerable load demand (i.e. $\max(u_i) = P^L_i \nu$). This restriction to the maximum LAA reflects the finite capacity of inactive loads the attacker can activate during an LAA. 
 
 
 \vspace{-0.2 cm}
 
 \subsection{Emergency Responses}
\label{sec:emer_res}
ERs refer to systems designed to protect sensitive power system components from excessive frequency deviations following a change in the active power balance. In this section we provide a brief discussion of ER employed, and refer the reader to \cite{goodridge2020} and \cite{goodridge2021} for a detailed mathematical description. 


\emph{ Generation shedding:} To protect synchronous generators, we model two independent schemes intended to disconnect the generator from the network: (i) RoCoF-induced generation shedding (RIGS) - the generator is disconnected when nodal RoCoF $|\ddot{\delta_i}|$ exceeds an upper threshold; (ii) over frequency generation shedding (OFGS) - generation is shed when nodal frequency $\dot{\delta_i}$ exceeds a pre-set upper limit. The binary variable $\psi_i$ models the activation of generation shedding at node $i$ during the simulation: $\psi_i = 1$ initially under normal operation; however, when either threshold is met and generator $i$ is disconnected, $\psi_i$ is set to 0 until the end of the simulation.

    
    
    
    \emph{ Under-frequency load shedding (UFLS):} We model UFLS schemes as a progressive disconnection of loads when the frequency $\dot{\delta}_i$ falls below a strictly decreasing sequence of four frequency thresholds $F^U \coloneqq \{F^U_1,\dots,F^U_4\}$ where $F^U_{j-1} > F^U_j$. In our model, at each frequency threshold, 10\% of equilibrium loads $P^L_i$ is automatically disconnected to arrest the decline in nodal frequency. Letting $\mathcal{R}_i \in \{0,1,2,3,4\}$ count the total number of UFLS activations at node $i$ at each time step $t$ in the simulation, the \textit{net load} is a dynamic variable in the power system model:
    
    {\small \begin{equation}
        \chi^L_i(\mathcal{R}_i) =  \bigg(1- 0.1\mathcal{R}_i \bigg)\bigg((1-\nu)P^L_i+  (1+\eta_i)\nu P^L_i\bigg)
    \end{equation}}
    
   \emph{ Line disconnection:} In the KTAS network, we model the disconnection of the line connecting Areas 1 and 2 when the power flow through the line, given by $\phi_{ij} \coloneqq B_{ij}E_i E_j sin(\delta_i - \delta_j)$, exceeds a pre-set power threshold $P^{\phi}$. When excess power flow is detected through the inter-connector line, the indicator $\Omega_{ij}$ switches from 1 to 0 for the remainder of the simulation, setting the $ij^{th}$ element of $B_{ij}$ to 0 \cite{goodridge2021}.

The ER model inspects the continuous time variables $\dot{\delta}_i(t)$, $\ddot{\delta}_i(t)$ and $\phi_{ij}(t)$ from the power system model~\eqref{eq:network_model} at regular time intervals. Once a criteria for activation is observed, the corresponding ER is activated. This is represented in \eqref{eq:network_model} as a discontinuity, where changes to the relevant input variables (power injection, load or network topology) are applied. Subsequently, the simulation is resumed with the new network parameters.


\section{Statistical Model for LAAs}\label{sc:statmodel}
In this section, we present the statistical model for the distribution of LAAs and describe the proposed rare-event sampling approach to identify the impactful LAAs. 
\subsection{Modeling the Unconditional Distribution of LAAs}
	

There are several studies that document the frequency and magnitude of cyber breaches in enterprise networks \cite{Edwards2016}. For instance, \cite{Edwards2016} demonstrates the size of data breaches can be well-modelled by the log-normal family of distributions. Assuming nodal LAAs magnitudes are independent and follow a similar distribution, we model $U\in \mathbb{R}^{N+L}  \coloneqq[|u_1|, \dots, |u_{N+L}| ]$ as
\begin{equation}
\label{eq:rho}
U \sim  \prod_{i=1}^{N+L} \text{Lognormal} (\mu_i,\sigma_i^2). 
\end{equation}
We also note that our analysis is not restricted to the log-normal family of distributions, and can be extended to any underlying distribution in a straightforward manner.

\vspace{-0.2 cm}

	 
	
%



	
	\subsection{Rare-Event Sampler for LAAs}
	\label{sc:skipping_sampler}
	If given an unconditional density $\rho$ over $\mathbb{R}^{N+L}$ and a rare event of interest $C \subset \mathbb{R}^{N+L}$, \textit{rare-event sampling} involves drawing elements from $\rho$ conditioned on the occurrence $C$. {The density of this conditional distribution for the element $U \in \mathbb{R}^{N+L}$ is:
	{ \begin{equation}
	    \pi(U) = \frac{\rho(U)\mathbbm{1}_C(U)}{\rho(C)},
	\label{eq:pi}
	\end{equation}}
	where $\rho(C)$ is the probability of the event $C$ occurring, and  
	\begin{equation}
	    \mathbbm{1}_C(U) =  \begin{cases} 1 & U \in C\\
	    0 & U \notin C. \end{cases}
	    \end{equation}
	\noindent In the context of our research, $C$ is the set of LAA magnitudes which result in the activation of at least one ER and $\rho$ is the distribution in \eqref{eq:rho}. As $C$ is expected to be rare, we employ the \textit{skipping sampler} MCMC algorithm to efficiently draw samples of $U \in C $.} The skipping sampler is formalised in Algorithm~\ref{alg:skip}. We provide an intuitive explanation of the algorithm in the following. 

	{\small \begin{algorithm}[!ht]
		\SetAlgoLined
		\SetKwInOut{Input}{Input}\SetKwInOut{Output}{Output}
		\textbf{Input:} initial state $U_1$;\\
		\textbf{for i = 1 to n do:}
		\BlankLine
		\DontPrintSemicolon
		Generate an initial proposal $\ensuremath{Z_1}$ distributed according to the density $q(\ensuremath{y} - \ensuremath{U_i})d\ensuremath{y}$;\\
		Calculate the direction $\ensuremath{\Phi}=\left( \ensuremath{Z_1}-\ensuremath{U_i}\right)/\left\Vert \ensuremath{Z_1}-\ensuremath{U_i}\right\Vert $;\\
		Generate a halting index $K \sim K_{\varphi}$;\\
		Set $k=1$ and:\\
		\While{$\ensuremath{Z}_k \notin C \textbf{\textup{ and }}$ $k< K$}{
			Generate a distance increment $R$ distributed according to $q_{r|\ensuremath{\Phi}}\left(r|\ensuremath{\Phi}\right)$;\\
			Set $\ensuremath{Z}_{k+1}=\ensuremath{Z}_k+ \ensuremath{\Phi} R$;\\
			k=k+1;\\
		}
		Set $\ensuremath{Z}\coloneqq \ensuremath{Z}_k$;\\
		Evaluate the acceptance probability:
		\begin{equation}\label{eq:ap2}
			\alpha(\ensuremath{U_i},\ensuremath{Z})=
			\begin{cases}
				\min\left(1, \frac{\pi(\ensuremath{Z})}{\pi(\ensuremath{U_i})} \right) & \text{ if } \pi(\ensuremath{U_i}) \neq 0, \\
				1, & \text{ otherwise, }
			\end{cases}
		\end{equation}
		Generate a uniform random variable $V$ on $(0,1)$;\\
		\uIf{$V \leq \alpha(\ensuremath{U_i},\ensuremath{Z})$}{$\ensuremath{U}_{i+1}=\ensuremath{Z}$;}
		\Else{$\ensuremath{U}_{i+1}=\ensuremath{U_i}$;}
		\KwRet{$\ensuremath{U}_{i+1}$}.\;
		\textbf{Output}: final sample $[U_1, \dots, U_h]$
		\caption{Skipping sampler algorithm}
		\label{alg:skip}
	\end{algorithm}}
	
As a Metropolis-class algorithm \cite{Moriarty2019}, the skipping sampler can be understood as a two step procedure: (1) a \textit{proposal step} where, starting from a state $U_n \in \mathbb{R}^{N+L}$, a potential new state $Z$ for the final sample is generated; (2) an \textit{acceptance/rejection} step, which determines whether the proposed state $Z$ is included in the final sample, according to a specified acceptance probability. This ensures the distribution of the sample follows the desired target distribution $\pi$. If it is accepted, the proposal is included in the final sample and becomes the starting state for the next proposal step. This procedure is repeated a desired number of times, after which the final sample is returned. 

The skipping sampler improves the sampling of $C$ by using a specialised proposal function which `skips' over $C^c$ - the set of LAAs which do not lead to the activation of an ER (which are not of interest); until the rare event is sampled or the skipping process is terminated. Thus, the proposal of the skipping sampler was designed to enable efficient transitions between connected components of $C$. Denoting the current state $U_n$, if the initial proposal $Z_1 \notin C$, we update the initial proposal (or `skip') by an adding an independent random distance increment $R_2$ in the direction $\Phi = \frac{Z_1 - U_n}{|Z_1 -U_n|}$, where $R_k$ has the conditional distribution of $\left\Vert Z_1-U_n\right\Vert$ conditioned on $\Phi$. The proposal function continues this linear update procedure until either $C$ is entered or the budget for skipping is exhausted \cite{Moriarty2019}. 
	
	%
	%
	\section{Simulations}\label{sec:Sims}
	


\subsection{Simulation Settings}
\label{sc:ktas}

\begin{figure}[tbph]
\vspace{-0.2 cm}
		\centering\includegraphics[width=2.5in]{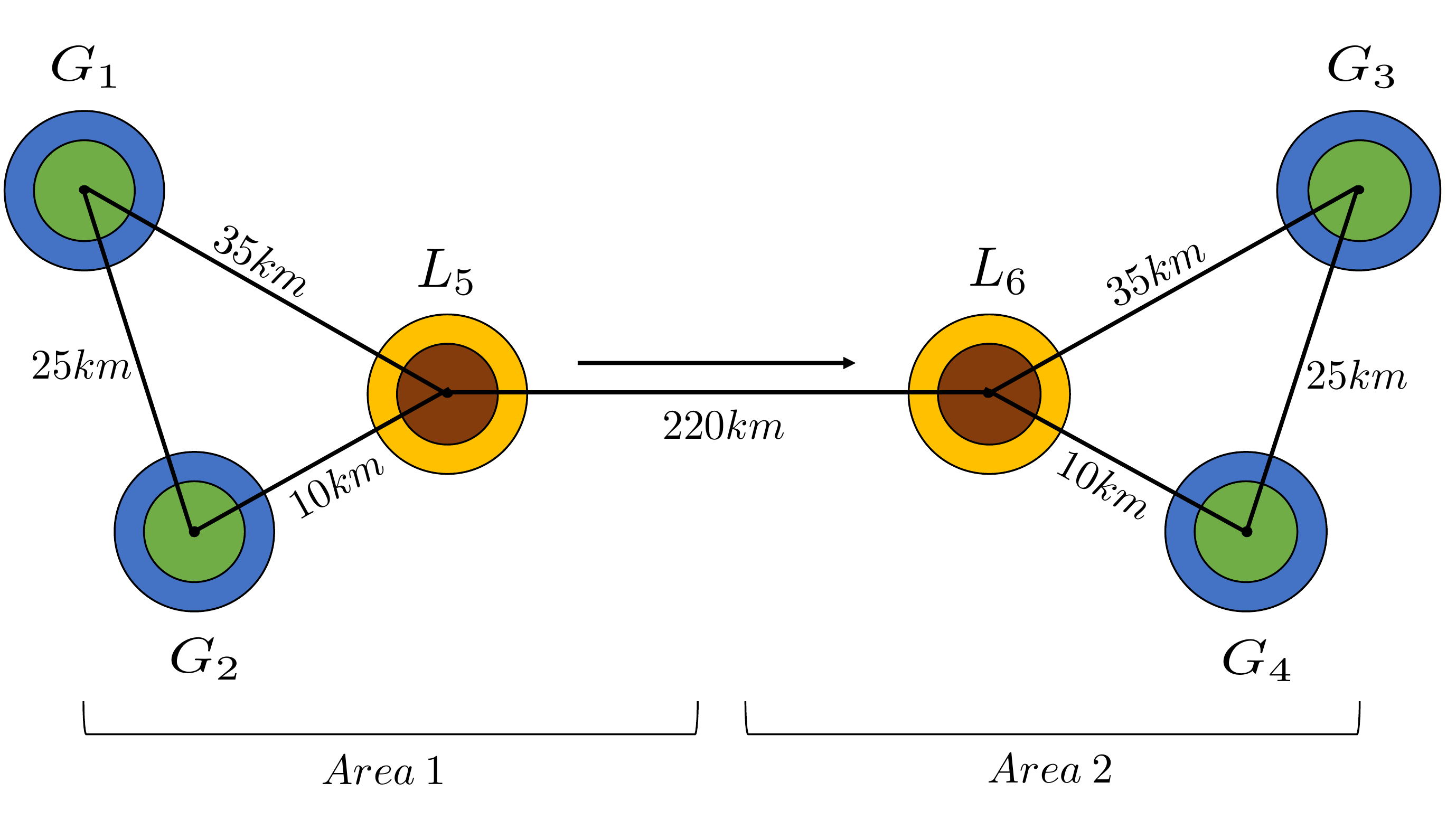}
		\caption{{Schematic drawing of the Kundur two-area 4 node network after Kron reduction. Generator buses (green circles) correspond to nodes $i=[1,\dots,4]$ and load buses (brown circles)  correspond to nodes $i=5,6$. Line lengths are indicated.}} 
		\label{fg:kundurbat}
		\vspace{-0.1 cm}
	\end{figure}

Our case study is based on the KTAS power grid \cite{kundur1994power}. We take a Kron reduced version consisting of $N=4$ generation buses and $L=2$ load buses  as shown in Figure \ref{fg:kundurbat}. At $t=0^-$, the system is modelled in equilibrium, with power flows from Area 1 to Area 2 through the line connecting nodes $5$ and $6$ (tie line). The system parameters are such that the system is $N-1$ secure, in the sense that the loss of a generator (in the absence of any other disturbance) does not trigger an ER. The initial conditions of the above system of equations, denoted $\delta_i\left(0\right)$, $E_i\left(0\right)$,  $\rho_i(0)$, $P_i^G$ and $P^L_i$ are set equal to equilibrium states which can be determined numerically, such that $\ddot{\delta}_i \approx 0$. These values along the those of the parameters of ~\eqref{eq:network_model} and \eqref{eq:network_model_load} can be found in the online Appendix \cite{extended}.
	
	
	
	%

To generate LAA instances, we implement the skipping sampler proposed in Section~\ref{sc:skipping_sampler}. During each proposal step, we sample an $N+L-$dimension LAA vector $U$ from Lognormal distribution as in \eqref{eq:rho}. We investigated various values for $\sigma_i \in [1, 8]$ which controls the rareness of large LAAs. For this study, we present the results for $\mu_i = 0$ and $\sigma_i =4$ for $i = 1,\dots, N+L$, and reserve a detailed sensitivity analysis for a dedicated manuscript. We apply $U$ as an input to the power system model~\eqref{eq:network_model} at $t = 0$, with frequency dynamics simulated for $15$ seconds following the LAA using MATLAB. We conduct $n=60,000$ proposals, which generated a final sample of $h \approx 10,500$ LAAs conditioned on the activation of at least one ER. This is an acceptance rate of 17.5\%, within the 15 - 48\% rate considered optimal for exploring of a sample \cite{Gamerman2006}. Following \cite{Moriarty2019}, no burn-in nor thinning was required, thus all samples collected were available for analysis. Since these responses are an  undesirable event (from a system operator's point of view), we label such instances as a ``network failure". 

\vspace{-0.2 cm}
\begin{figure}[!ht]
\centering
\includegraphics[width=0.5\textwidth]{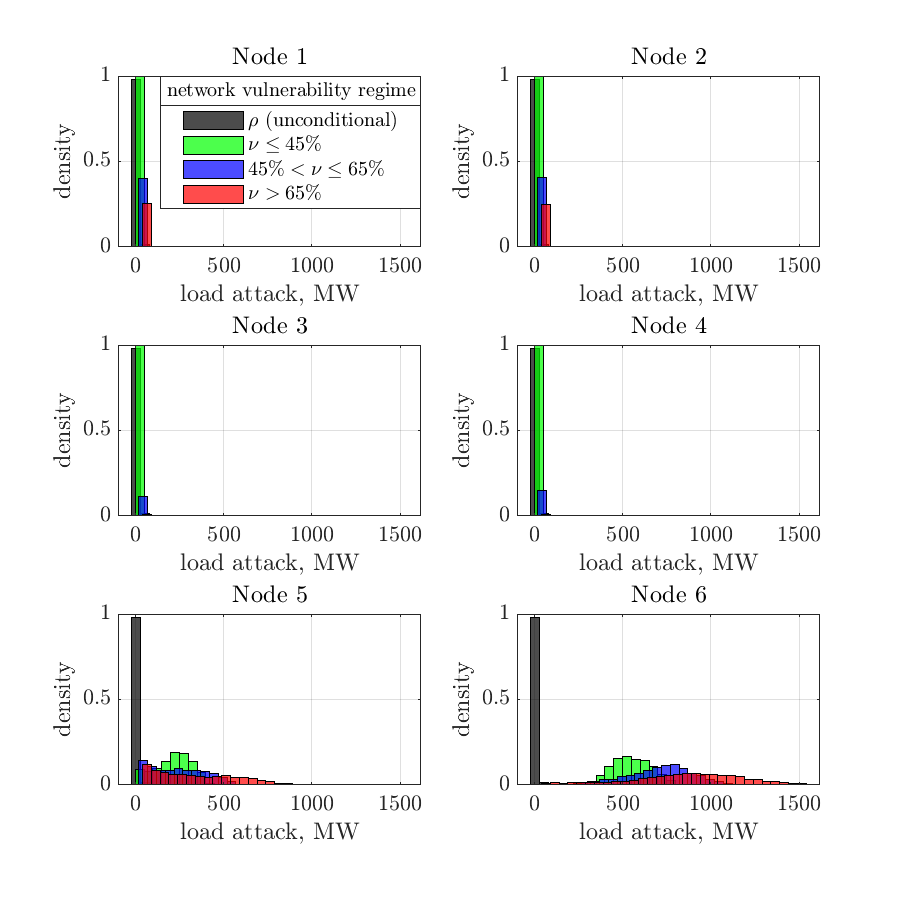}
\caption{{Distribution of the absolute value LAAs at each node of the KTAS network conditioned on a network failure event, for different levels of network vulnerability $\nu$ and the unconditional distribution ($\rho$). Note - bins for successive values of $\nu$ are offset slightly to the right to improve readability.}}
\label{fig:LAA_Distribution}
\vspace{-0.1 cm}
\end{figure}


\subsection{Simulation Results}
We evaluate the susceptibility of the network to LAAs at both a local and global scope for three regimes of network vulnerability to LAAs- a `secure network' ($\nu \le 45\%)$, a `moderately vulnerable network' ($45\% < \nu \le 65\%)$ and a `highly vulnerable network' ($\nu > 65\%)$. 

{\bf Local Analysis -- Identifying the vulnerable nodes:}
Figure~\ref{fig:LAA_Distribution} plots the distribution of LAA magnitudes at each node, conditioned on the occurrence of a network failure, for different degrees of network vulnerability $\nu$. We observe the following : (i) 
only at nodes $5$ and $6$ does the conditional distribution of LAA magnitudes differ significantly from $\rho$, with network failures associated with larger magnitude LAAs, located in the low density region of $\rho$. This is driven by the design of the KTAS network, where most loads are concentrated at nodes $5$ and $6$, giving the attacker sufficient leverage over system frequency to trigger a network failure. Thus, network failures are primarily driven by LAAs at these nodes, mostly independent of the LAA magnitude at nodes $1-4$; (ii) as network vulnerability $\nu$ increases, the distribution of LAA magnitudes at node 6 shifts rightwards, implying larger magnitude LAAs at node $6$ become more prevalent in the sample; however (iii) we note that large magnitude LAAs are rare under their assumed Log-normal distribution. Thus, Fig.~\ref{fig:LAA_Distribution} reveals network failures, driven by large LAA magnitudes at node $6$, are indeed low probability events. Summarily, these results suggest nodes~$5$ and $6$ are the critical nodes in the KTAS network from which the attacker can launch the most impactful attacks, regardless of the degree of network vulnerability.

\begin{figure}[!h]
\centering
\includegraphics[width=0.5\textwidth]{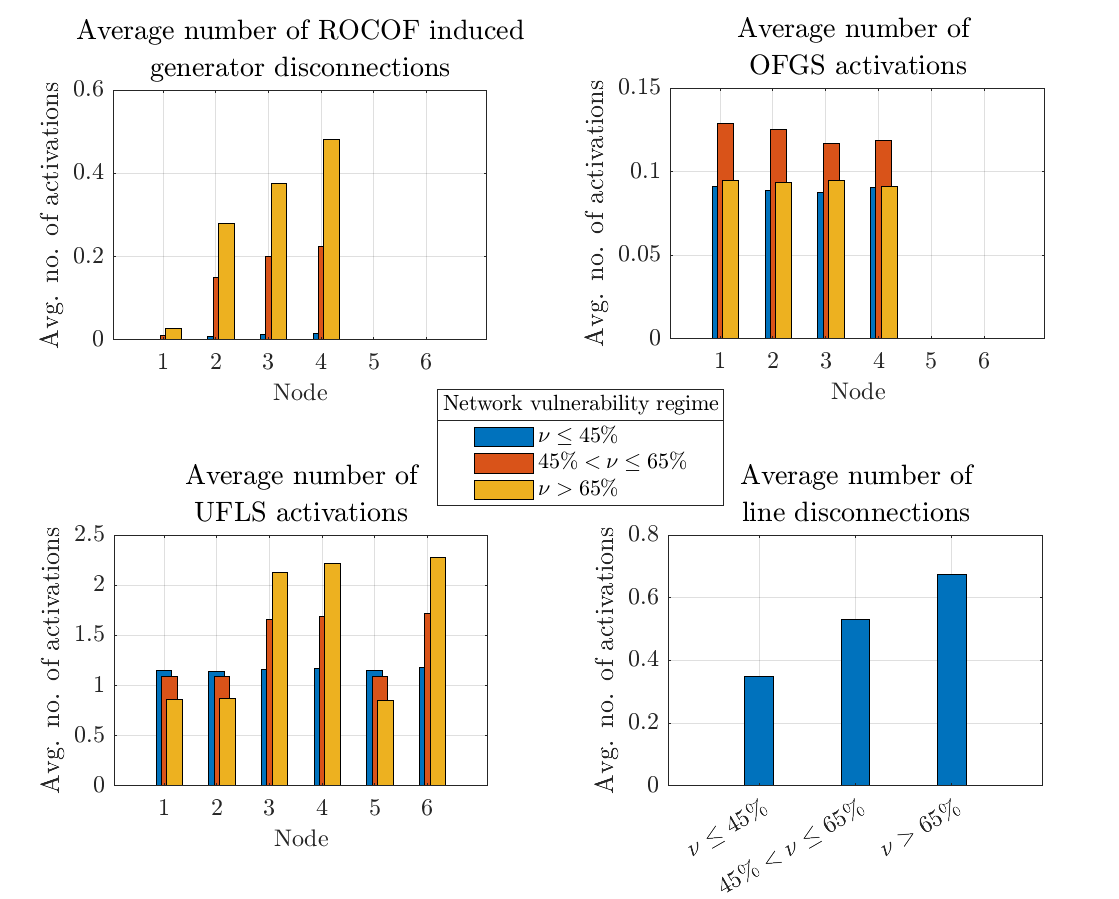}
\caption{{Average number of each ER per sample $\epsilon_i, i = 1,\dots,4$ (defined in \eqref{eqn:epsilon_def}) in the KTAS network for different degrees of network vulnerability to LAAs.}}
\label{fig:nodal_Failure}
\vspace{-0.1 cm}
\end{figure}


Figure~\ref{fig:nodal_Failure} illustrates the average number of activations of each ER per sample for different levels of network vulnerability, defined as 
 {\small\begin{align}
     \epsilon_{f} = \frac{\text{Total ER$_f$ activations in the sample}}{\text{Sample size} }. \label{eqn:epsilon_def}
 \end{align}}
 Herein, the indices $f = 1,\dots,4$ correspond to 
 RoCoF-induced generation shedding (RIGS), over-frequency generator shedding (OFGS), UFLS activations, and inter-connector line trips respectively. Note that $\epsilon_{f} \in [0,1]$ for $f = 1,2,4$. However, $\epsilon_3 \in [0,4]$, as each node can experience a maximum of 4 UFLS events ($f=3$) during the simulation (see Section~\ref{sec:emer_res}). 
 
 For secure networks, i.e, when $\nu \le 45\%$, the average number of RIGS is comparatively small, as low-magnitude LAAs are unable to induce sufficiently large RoCoF deviations to trigger generator disconnection. Additionally, we observe all generator nodes experience OFGS and UFLS events at similar rates of $\epsilon_2 \approx 0.1$ and $\epsilon_3 \approx 1$ respectively. Together, these imply for a secure KTAS network, each class of ER is similarly likely at any node in the network. Thus, to counter the potential threat of LAAs in a secure KTAS network will require a system-wide solution, e.g. - a coordinated automatic generation control (AGC) system.

As $\nu$ increases, it grants the attacker greater authority over network loads and the ability to significantly disrupt the active power balance of the network. This is associated with increased rates of RIGS, UFLS and line disconnections, as these are triggered by large power deviations. However, when $\nu > 65\%$, the average number of OFGS responses declines, as large changes in loads result in RIGS dominating generation shedding events. 
In contrast to secure networks, where nodes are similarly vulnerable to UFLS events under LAAs, Figure~\ref{fig:nodal_Failure} reveals a differential in the nodal risk of an UFLS event in moderately and highly vulnerable networks. For example, when $\nu > 65\%$, nodes $3, 4$ and $6$ experience approximately twice as many UFLS events as nodes $1, 2$ and $5$. This can be best understood through a global analysis of the KTAS network, which is discussed next. 

{\bf Global Analysis - Spatial Distribution of LAAs:}
Recall that the KTAS network is comprised of two areas- Area 1 (nodes 1, 2 and 5) with excess generation, and Area 2 (nodes 3, 4 and 6) with excess demand. With sufficient authority over network loads, an attacker can exploit the global pre-LAA power imbalances of each area to trigger an ER by decreasing loads in Area 1 to exacerbate the excess generation, and increasing loads in Area 2 to exacerbate the generation deficit. Thus, as $\nu$ increases, we observe fewer UFLS responses in Area 1, more UFLS responses in Area 2, and a substantial increase in power transferred across the inter-connector, increasing the rates of line disconnections (Figure~\ref{fig:nodal_Failure}).

\begin{figure}[!t]
\hspace{-1cm}
\includegraphics[width=0.55\textwidth]{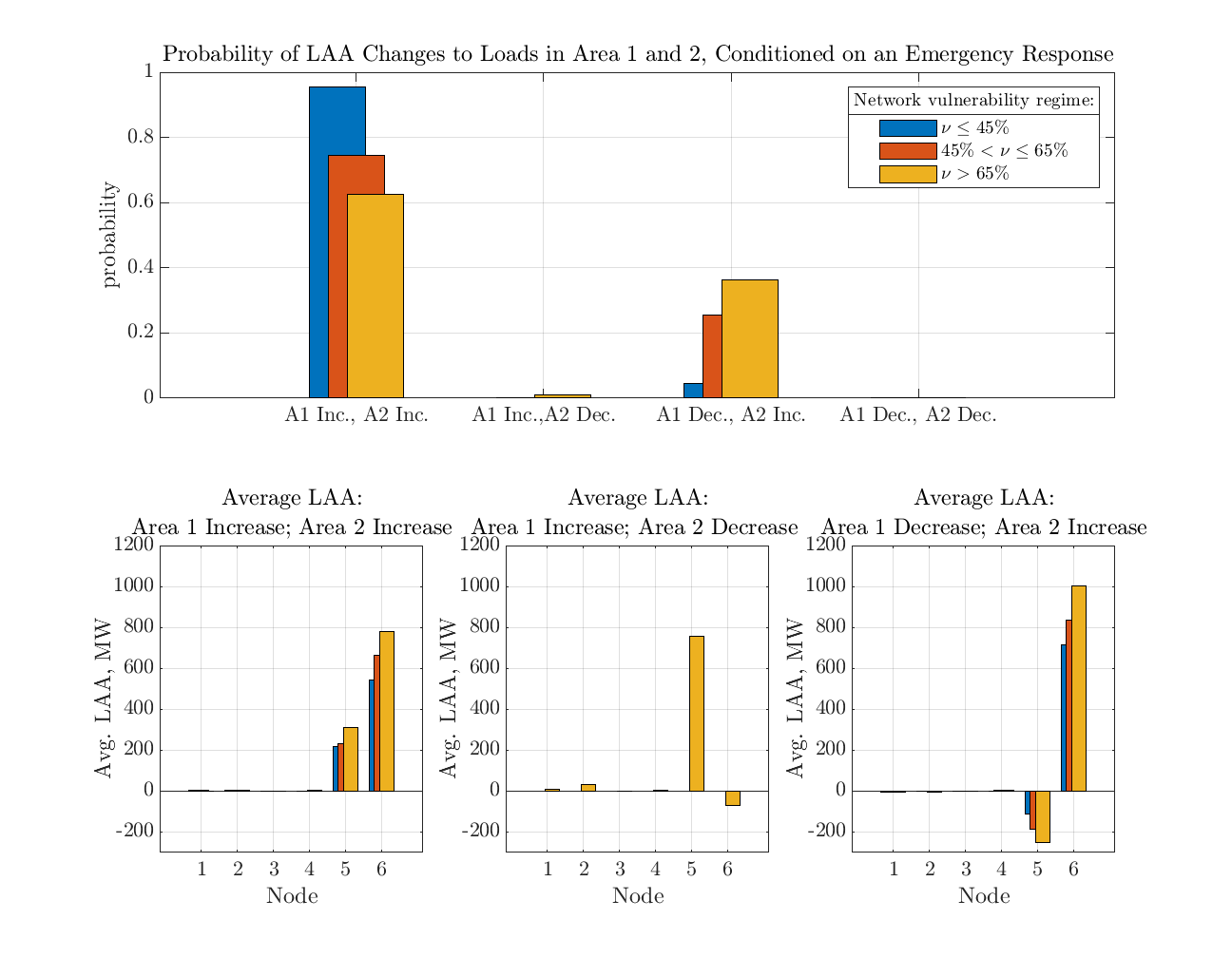}
\caption{{Probability of network failure for increase/decrease of system load in each area for the KTAS network.}}
\label{fig:Area_Failure}
\vspace{-0.1 cm}
\end{figure}

The global susceptibility of the KTAS network to LAAs is also illustrated by Figure~\ref{fig:Area_Failure}, which plots the probability of the relative changes in loads in each Area of the KTAS network, conditioned on the occurrence of a network failure. First, we note the KTAS network is secure against global reduction of loads in both Areas, as it is likely governors are able to adjust generator output to handle the loss of loads in all vulnerability regimes. Instead, we observe that failure events occur primarily for large increases in loads, dominated by two scenarios: (i) when the LAA increases the demand in both Areas- this scenario is the primary driver of failure events when KTAS is secure, as the attacker increases loads at both nodes 5 and 6 beyond the maximum power capabilities of generators, triggering a failure anywhere in the network; (ii) the attacker decreases loads in Area~1 and increases them in Area~2- this scenario is rare when the KTAS is secure, as the attacker generally lacks sufficient authority to exploit the imbalances of the KTAS network. However, as $\nu$ (and the attacker's authority) increases, so too does the conditional probability of scenario (ii) (see Figure~\ref{fig:nodal_Failure}). 

\section{Conclusions and Future Research}
\label{sec:Conc}
In this work, we present a framework to evaluate the impact of LAAs on power grid operations using a rare-event sampling approach. 
The proposed approach provides a comprehensive framework to examine the impact of LAAs under different potential spatial distributions of LAAs across the power network. Our results identify the nodes from which the attacker can launch the most impactful LAA and further illustrate how the attacker can exploit the inter-area power imbalances in the network to trigger ER events. 
Future work includes (i) an extension of the analysis to dynamic LAAs against power grids \cite{AminiLAA2018}, (ii) considering correlations in LAA injections, (iii) showing scalability of the proposed method to large-dimensional systems.


\bibliographystyle{IEEEtran}
\bibliography{IEEEabrv,bibliography}

\newpage

\end{document}